\documentclass[pdflatex,sn-mathphys-num]{sn-jnl}

\usepackage{graphicx}%
\usepackage{multirow}%
\usepackage{amsmath,amssymb,amsfonts}%
\usepackage{amsthm}%
\usepackage{mathrsfs}%
\usepackage[title]{appendix}%
\usepackage{xcolor}%
\usepackage{textcomp}%
\usepackage{manyfoot}%
\usepackage{booktabs}%
\usepackage{algorithm}%
\usepackage{algorithmicx}%
\usepackage{algpseudocode}%
\usepackage{listings}%
\usepackage{bm}
\usepackage{hyperref}

\newcommand{\bra}[1]{\langle #1|}
\newcommand{\ket}[1]{|#1\rangle}

\begin{document}

\title[Post-Processing Reconstruction of Finite-Temperature Dynamics]{Gate-Based Quantum Simulation of Finite-Temperature Open-System Dynamics: A Post-Processing Reconstruction Approach for Collision Models}

\author*[1]{\fnm{Kazunari} \sur{Hashimoto}}\email{hashimotok@yamanashi.ac.jp}
\equalcont{These authors contributed equally to this work.}

\author[1]{\fnm{Chikako} \sur{Uchiyama}}\email{hchikako@yamanashi.ac.jp}
\equalcont{These authors contributed equally to this work.}

\affil*[1]{\orgdiv{Faculty of Engineering}, \orgname{University of Yamanashi}, \orgaddress{\street{4-3-11 Takeda}, \city{Kofu}, \postcode{400-8511}, \country{Japan}}}

\maketitle

\begin{abstract}

Simulating open-system dynamics on gate-based quantum computers offers a route to exploring dissipative quantum phenomena through programmable quantum processing.
Collision models provide a direct and flexible framework by representing the environment through sequential system--ancilla collisions.
At finite temperatures, however, they typically require repeated preparation of thermal ancillas, adding circuit overhead and gate errors.
Here, we show that, for a broad class of $U(1)$-symmetric system--ancilla interactions, finite-temperature dynamics can instead be reconstructed from simulations with pure ancillas in orthogonal basis states.
The target temperature enters only through classical post-processing of the resulting pure-branch data, allowing the same data to be reused for different temperatures without additional quantum processing.
Bypassing explicit thermal-state preparation reduces ancilla-preparation overhead and preparation-induced errors.
\end{abstract}

\section{Introduction}
\label{sec:introduction}

Simulating open quantum systems on gate-based quantum computers offers a programmable route to exploring dissipative quantum phenomena~\cite{schlimgen2021,kamakari2022,liu2025}.
A variety of quantum algorithms have been developed for this purpose, particularly for Lindblad dynamics~\cite{kliesch2011,barthel2012,cleve2016,childs2016,ding2025lower,schlimgen2022,li2022,ding2024}.
These approaches employ different algorithmic constructions to represent nonunitary open-system evolution using quantum circuits.
Alongside these approaches, collision models (also known as repeated-interaction models) provide a direct and flexible framework in which the environment is represented through sequential system--ancilla interactions~\cite{rau1963,ziman2002,scarani2002,ziman2005all,ziman2005desc,attal2006,chimonidou08,giovannetti2012JPhysB,giovannetti2012PRL,ciccarello2013PRA,ciccarello2013PS,bruneau2014,lorenzo2016,ciccarello2017QM,guarnieri2020,campbell2021,ciccarello2022PR,cusumano2022entropy,cattaneo2022brief}.
Their simple and modular structure maps naturally onto gate-based quantum circuits~\cite{garcia2020ibm,cattaneo2021,melo2022,cattaneo2023,dibartolomeo2024,borras2025,garg25,pocrnic2025}, while retaining a microscopic description in terms of system--environment interactions and recovering Lindblad dynamics in appropriate continuous-time limits.

At finite temperatures, however, implementing collision models typically requires repeated preparation of thermal ancilla states corresponding to the desired environmental temperature.
On gate-based quantum computers, such mixed states are commonly generated through purification, in which the thermal ancilla is realized as the reduced state of a pure entangled state formed with an additional qubit and prepared using a temperature-dependent single-qubit rotation and an entangling gate~\cite{pocrnic2025,takahashi1996,zhu2020PNAS,chowdhury2020arxiv}.
Repeating this preparation throughout the simulation increases circuit overhead and introduces additional gate errors.
More fundamentally, because the ancilla preparation depends explicitly on the temperature of the simulated environment, the quantum processing itself becomes temperature dependent, requiring a new quantum-processing step for each temperature.

In this work, we avoid temperature-dependent ancilla preparation by using only two orthogonal pure ancilla states and incorporating the environmental temperature afterward by classically combining and correcting the resulting pure-branch data.
We refer to this procedure as post-processing reconstruction.
This reconstruction is nontrivial under repeated collisions: although the one-step thermal collision map is a statistical mixture of the two pure-branch maps, its repeated application does not generally reduce to the same mixture of the corresponding repeated pure-branch dynamics.
For the broad class of system--ancilla interactions with \(U(1)\) symmetry, the resulting symmetry-induced structure resolves this difficulty by separating the dynamics into population and coherence sectors and, in the coherence sector, by enforcing a common contraction factor for the two pure branches while allowing branch-dependent phases.
These properties lead to distinct reconstruction rules for the two sectors.
Population dynamics is recovered exactly as a convex mixture of the two pure-branch dynamics, whereas coherence is reconstructed from the dynamics of a single pure branch using an analytically determined contraction factor and phase correction.
Because the underlying pure-branch data are temperature independent, the same data can be reused to retarget the dynamics to different temperatures without additional quantum processing.
Finally, comparison with purification-based implementations shows that eliminating explicit thermal-state preparation reduces the associated ancilla-preparation overhead and preparation-induced errors.

The paper is organized as follows.
In Sec.~\ref{sec:model}, we introduce the collision model and derive the symmetry-induced sector decomposition.
In Sec.~\ref{sec:postprocessing}, we present the post-processing reconstruction formulae for the population and coherence sectors.
In Sec.~\ref{sec:errorsec}, we compare the ancilla-preparation overhead of the two implementations and analyze the resulting preparation-induced errors within the broader error decomposition of repeated-interaction simulations.
In Sec.~\ref{sec:discussion}, we discuss the physical scope, dynamical conditions, and
limitations of the proposed reconstruction framework.
Section~\ref{sec:conclusion} concludes the paper.

\section{Collision model and symmetry constraints}
\label{sec:model}
In this section, we introduce the finite-temperature collision model and establish the symmetry-induced dynamical structure underlying the post-processing reconstruction developed in Sec.~\ref{sec:postprocessing}.
We first define the thermal and pure-branch collision maps and then show how the \(U(1)\) symmetry constrains their population and coherence dynamics.

\subsection{Collision model with thermal ancillas}
\label{subsec:collision_model}
We consider a collision model in which a system qubit \(S\) interacts sequentially with a stream of ancilla qubits \(A\).
Each collision is generated by a unitary operator
\begin{equation}
U:=e^{-iH\tau},
\end{equation}
where \(H\) is the system--ancilla Hamiltonian and \(\tau\) is the duration of the collision.
After each collision, the ancilla is discarded and replaced by a fresh ancilla.
For an ancilla state \(\rho_A\), the one-step collision map is
\begin{equation}
\Phi[\rho_S]
:=
\mathrm{Tr}_A\!\left[
U(\rho_S\otimes\rho_A)U^\dagger
\right].
\label{eq:collision_map}
\end{equation}
The map is completely positive and trace preserving (CPTP)~\cite{ciccarello2022PR}.

Each ancilla is prepared in the same initial state $\rho_A$.
At finite temperature, the ancilla is prepared in the Gibbs state
\begin{equation}
\rho_A^{\rm th}
:=
p_\uparrow \ket{\!\uparrow}\!\bra{\uparrow\!}
+
p_\downarrow \ket{\!\downarrow}\!\bra{\downarrow\!},
\qquad
p_\uparrow+p_\downarrow=1,
\label{eq:thermal_ancilla}
\end{equation}
where \(|\uparrow\rangle\) and \(|\downarrow\rangle\) are eigenstates
of \(\sigma_z^A\), respectively, and \(p_\uparrow\) and
\(p_\downarrow\) are the corresponding thermal
probabilities determined by the target temperature \(T\) through the
Gibbs distribution.
The corresponding finite-temperature collision map is defined as
\begin{equation}
\Phi_{{\rm th}}[\rho_S]
:=
\mathrm{Tr}_A\!\left[
U(\rho_S\otimes\rho_A^{\rm th})U^\dagger
\right].
\label{eq:cm_thermal}
\end{equation}
We also define the collision maps associated with pure ancillas as
\begin{equation}
\Phi_s[\rho_S]
:=
\mathrm{Tr}_A\!\left[
U(\rho_S\otimes |s\rangle\langle s|)U^\dagger
\right],
\qquad
s=\uparrow,\downarrow.
\label{eq:pure_branch_maps}
\end{equation}
We refer to the two maps $\Phi_\uparrow$ and $\Phi_\downarrow$ as the pure branches.

Since the collision map \(\Phi_{\rm th}\) is linear on the ancilla state and the thermal state \(\rho_A^{\rm{th}}\) is a convex mixture of the pure states,
the one-step finite-temperature collision map is given by the convex mixture of the pure branches,
\begin{equation}
\Phi_{\rm th}
=
p_\uparrow\Phi_\uparrow+p_\downarrow\Phi_\downarrow.
\label{eq:one_step_thermal_channel}
\end{equation}
The central question is whether this one-step convex relation remains valid after repeated applications of $\Phi_{\rm th}$.

\subsection{\texorpdfstring{$U(1)$}{U(1)} symmetry and sector decomposition}
\label{subsec:u1_sector}

We focus on system--ancilla interactions with a \(U(1)\) symmetry under simultaneous rotations of the system and ancilla about the quantization axis \(z\).
This symmetry is central to the reconstruction developed below.
It separates population dynamics along the \(z\) axis from coherence dynamics in the transverse plane, allowing the two sectors to be treated independently.
As shown in Sec.~\ref{subsec:pure_branch_maps}, it further constrains the two pure branches to share a common contraction factor in the coherence sector, with their branch dependence appearing only through the phase.
Both properties are essential to the reconstruction derived in Sec.~\ref{sec:postprocessing}.

At the Hamiltonian level, this symmetry is expressed by the commutation relation
\begin{equation}
[H,\sigma_z^S+\sigma_z^A]=0.
\label{eq:u1_symmetry}
\end{equation}
For the two-qubit system considered here, this condition is equivalent to conservation of the total excitation number during each collision.

To see how the \(U(1)\) symmetry constrains the collision dynamics,
we represent the system state in terms of the Bloch vector,
\begin{equation}
\rho_S
=
\frac{1}{2}
\left(
I+\bm r\cdot\bm\sigma
\right),
\qquad
\bm r=(x,y,z)^T,
\end{equation}
where \({\bm\sigma}=(\sigma_x^S,\sigma_y^S,\sigma_z^S)\) denotes the Pauli operators for the system.
In this representation, any single-qubit CPTP map acts as an affine transformation,
\begin{equation}
\bm r\mapsto T\bm r+\bm t,
\label{eq:bloch_affine}
\end{equation}
where $T$ is a real $3\times3$ matrix describing the linear part of the map, and $\bm t$ represents an affine shift.
The Bloch representation makes the dynamical consequences of the $U(1)$ symmetry particularly transparent.

As shown explicitly in Appendix~A, the symmetry condition in Eq.~\eqref{eq:u1_symmetry}, together with an ancilla state diagonal in the $\sigma_z$ basis, makes the reduced collision map covariant under rotations about the $z$ axis.
This covariance prevents mixing between the longitudinal ($z$) and transverse ($x,y$) components and constrains the affine map to the form
\begin{equation}
T=
\begin{pmatrix}
\mu & \nu & 0\\
-\nu & \mu & 0\\
0 & 0 & \gamma
\end{pmatrix},
\qquad
\bm t=
\begin{pmatrix}
0\\
0\\
f
\end{pmatrix}.
\label{eq:u1_bloch_form}
\end{equation}
Thus, population and coherence dynamics form two independent sectors.
The transverse block can be written as
\begin{equation}\label{eq:cont_rot}
\begin{pmatrix}
\mu & \nu\\
-\nu & \mu
\end{pmatrix}
=\sqrt{\mu^2+\nu^2}
\begin{pmatrix}
\cos\theta & \sin\theta\\
-\sin\theta & \cos\theta
\end{pmatrix},
\qquad
\theta:=\arctan\frac{\nu}{\mu},
\end{equation}
corresponding to a contraction accompanied by a rotation in the $(x,y)$ plane, while the longitudinal component describes population relaxation.

\subsection{Pure-branch collision maps}
\label{subsec:pure_branch_maps}

The $U(1)$ symmetry established above separates the collision dynamics into independent population and coherence sectors.
We now determine how this symmetry-constrained structure is realized for the two pure ancilla states that constitute the thermal ancilla.
As shown explicitly in Appendix~\ref{app:pure_branch_maps}, the two pure branches share the same dissipative contraction, while their dependence on the ancilla state appears through the affine shift in the population sector and the accumulated phase in the coherence sector.
These properties will provide the building blocks for reconstructing the finite-temperature dynamics in Sec.~\ref{sec:postprocessing}.

Let us first consider the population sector.
For the pure branches \(s=\uparrow,\downarrow\), the population dynamics takes the affine form
\begin{equation}
\Phi_s:
\quad
z
\mapsto
\gamma z
+
(1-\gamma)z_s,
\label{eq:pure_population_map}
\end{equation}
where
\[
z_\uparrow=+1,
\qquad
z_\downarrow=-1,
\]
and
\(
0\le\gamma\le1
\)
is the common relaxation coefficient shared by both branches.
The explicit derivation from the most general \(U(1)\)-symmetric two-qubit unitary is given in Appendix~B.1.

Repeated application yields
\begin{equation}
z_n^{(s)}
=
\gamma^n z_0
+
(1-\gamma^n)z_s,
\label{eq:gensolpurez}
\end{equation}
where \(z_0\) is the initial system polarization.
Thus, both branches relax with the same rate toward different fixed points determined by the corresponding ancilla state.

We next consider the coherence sector.
For convenience, we introduce the complex coherence variable
\[
w:=x+iy.
\]
As shown in Sec.~\ref{subsec:u1_sector}, the transverse dynamics is described by a contraction accompanied by a rotation in the \((x,y)\) plane.
For each pure branch, this action reduces to a simple complex multiplication,
\begin{equation}
\Phi_\uparrow:
\quad
w\mapsto\lambda_\uparrow w,
\qquad
\Phi_\downarrow:
\quad
w\mapsto\lambda_\downarrow w,
\end{equation}
with
\begin{equation}
\lambda_s
=
\sqrt{\gamma}\,
e^{i\theta_s},
\qquad
s=\uparrow,\downarrow.
\label{eq:lambda_pure}
\end{equation}
The contraction factor
\(\sqrt{\gamma}\)
is common to the two branches, whereas the phases
\(\theta_\uparrow\)
and
\(\theta_\downarrow\)
depend on the branch.
Their explicit expressions are derived in Appendix~B.2.

Repeated application gives
\begin{equation}
w_n^{(s)}
=
\gamma^{n/2}
e^{in\theta_s}
w_0,
\qquad
s=\uparrow,\downarrow,
\label{eq:gen_sol_coh}
\end{equation}
showing that the coherence amplitude decays at the same rate in both branches, while the accumulated phase remains branch dependent.

The pure-branch dynamics is therefore characterized by a common dissipative contraction governed by \(\gamma\), with the branch dependence confined to the population fixed point and the coherence phase.
This structure forms the basis of the post-processing reconstruction developed in the next section.

\section{Post-processing reconstruction of finite-temperature dynamics}
\label{sec:postprocessing}
Although a thermal ancilla state is a statistical mixture of the two pure basis states, it does not generally follow that the dynamics after repeated collisions is given by the same mixture of the corresponding pure-branch dynamics, because repeated application of the thermal collision map can generate contributions involving different branches.
The symmetry-induced structure derived above makes these repeated-collision contributions tractable by separating the population and coherence sectors, for which distinct reconstruction rules can be established.
We therefore treat the two sectors separately below.

\subsection{Population sector: exact convex reconstruction}
\label{subsec:post_population}

Let us first examine the population sector.
Compared with the pure branches, the thermal ancilla changes only the fixed point of the population dynamics, while the relaxation coefficient \(\gamma\) remains unchanged.
Indeed, applying the same calculation as in the pure-ancilla case to thermal ancillas, we obtain the one-step population dynamics
\begin{equation}
\Phi_{\rm th}:z\mapsto\gamma z+(1-\gamma)z_A,
\end{equation}
where \(z_A:=p_\uparrow-p_\downarrow\) is the \(z\) component of the
thermal ancilla's initial Bloch vector and is determined by the target
temperature through the thermal probabilities \(p_\uparrow\) and
\(p_\downarrow\).
Repeated application of the sequential relation provides the general solution
\begin{equation}
z_n^{\rm th}
=
\gamma^n z_0
+
(1-\gamma^n)z_A.
\label{eq:zn_thermal}
\end{equation}
It ensures the asymptotic relaxation of the system toward the polarization of the thermal ancilla,
\begin{equation}
\lim_{n\to\infty}z_n^{\rm th}=z_{A}.
\end{equation}

As noted above, the fact that the thermal ancilla is a statistical mixture does not by itself guarantee that the same convex-mixture relation is preserved after repeated collisions.
For the population sector, however, the common relaxation coefficient of the two pure branches makes such an exact reconstruction possible.
Using Eq.~\eqref{eq:gensolpurez} together with
\(z_A=p_{\uparrow}-p_{\downarrow}\), we find
\begin{align}
p_\uparrow z_n^{(\uparrow)}
+
p_\downarrow z_n^{(\downarrow)}
&=
p_\uparrow
\left[
\gamma^n z_0+(1-\gamma^n)
\right]
+
p_\downarrow
\left[
\gamma^n z_0-(1-\gamma^n)
\right]
\nonumber\\
&=
\gamma^n z_0
+
(1-\gamma^n)(p_\uparrow-p_\downarrow)
\nonumber\\
&=
z_n^{\rm th}.
\end{align}
Therefore,
\begin{equation}
z_n^{\rm th}
=
p_\uparrow z_n^{(\uparrow)}
+
p_\downarrow z_n^{(\downarrow)}.
\label{eq:population_exact_reconstruction}
\end{equation}
Importantly, Eq.~\eqref{eq:population_exact_reconstruction} holds for arbitrary collision number \(n\): although repeated application of a mixture of collision maps does not generally reduce to the same mixture of the repeatedly applied pure-branch maps, the population sector retains this convex-mixture structure exactly.
Thus, the finite-temperature population dynamics can be reconstructed from the two pure-branch data sets using only the thermal probabilities as classical weights.

\subsection{Coherence sector: phase averaging and correction}
\label{subsec:post_coherence}

We next turn to the coherence sector, where the reconstruction differs essentially from the population case.
Although the two pure branches share the same contraction factor, they acquire different phases at each collision.
Consequently, the convex-mixture structure found in Eq.~\eqref{eq:population_exact_reconstruction} is not preserved for the $n$-step coherence dynamics.
Instead, the branch-dependent phases must first be combined at the level of the one-step finite-temperature collision map and then propagated through successive collisions.

Using Eq.~\eqref{eq:lambda_pure}, the finite-temperature collision map
acting on the coherence variable \(w\) can be written as
\begin{equation}
\Phi_{\rm th}:w\mapsto\lambda_{\rm th}w,
\end{equation}
where
\begin{align}
\lambda_{\rm th}
&:=
p_\uparrow\lambda_\uparrow
+
p_\downarrow\lambda_\downarrow
\nonumber\\
&=
\sqrt{\gamma}
\left(
p_\uparrow e^{i\theta_\uparrow}
+
p_\downarrow e^{i\theta_\downarrow}
\right).
\label{eq:lambda_th}
\end{align}
We therefore define
\begin{equation}
q e^{i\Theta}:=
p_\uparrow e^{i\theta_\uparrow}
+
p_\downarrow e^{i\theta_\downarrow},
\label{eq:qTheta_def}
\end{equation}
where
\begin{equation}
q:=
\left|
p_\uparrow e^{i\theta_\uparrow}
+
p_\downarrow e^{i\theta_\downarrow}
\right|,
\qquad
\Theta:=
\arg\left(
p_\uparrow e^{i\theta_\uparrow}
+
p_\downarrow e^{i\theta_\downarrow}
\right).
\end{equation}
We thus have
\begin{equation}
\lambda_{\rm th}
=
\sqrt{\gamma}\,q e^{i\Theta}.
\end{equation}
The factor \(q\) is explicitly
\begin{equation}
q=
\sqrt{
p_\uparrow^2+p_\downarrow^2
+
2p_\uparrow p_\downarrow
\cos(\theta_\uparrow-\theta_\downarrow)
},
\label{eq:q_formula}
\end{equation}
which satisfies \(0\leq q\leq1\).
It represents the additional coherence contraction arising from statistical averaging over the branch-dependent phases, while \(\Theta\) gives the corresponding effective phase rotation.

Repeated application of the finite-temperature collision map then gives
\begin{equation}
w_n^{\rm th}
=
\gamma^{n/2}q^n e^{in\Theta}w_0.
\label{eq:w_th_n}
\end{equation}
This result also makes explicit why a direct convex mixture of the $n$-step pure-branch coherences does not generally reproduce the finite-temperature coherence dynamics: the latter contains the factor
\(
q^n e^{in\Theta}
=
\left(
p_\uparrow e^{i\theta_\uparrow}
+
p_\downarrow e^{i\theta_\downarrow}
\right)^n,
\)
rather than
\(
p_\uparrow e^{in\theta_\uparrow}
+
p_\downarrow e^{in\theta_\downarrow}.
\)

Nevertheless, the symmetry-induced common contraction factor of the two pure branches provides the key structure that makes single-branch reconstruction possible. Because both branches share the same factor \(\gamma^{n/2}\) and differ only in their accumulated phases, Eq.~\eqref{eq:w_th_n} can be related directly to the coherence dynamics of either pure branch in Eq.~\eqref{eq:gen_sol_coh}.
The finite-temperature coherence dynamics can therefore be reconstructed from the \(\uparrow\) branch as
\begin{equation}
w_n^{\rm th}
=
q^n e^{in(\Theta-\theta_\uparrow)}
w_n^{(\uparrow)},
\label{eq:coh_recon_up}
\end{equation}
or equivalently from the \(\downarrow\) branch as
\begin{equation}
w_n^{\rm th}
=
q^n e^{in(\Theta-\theta_\downarrow)}
w_n^{(\downarrow)}.
\label{eq:coh_recon_down}
\end{equation}
Thus, unlike the population reconstruction, which requires the convex combination of both pure branches, the finite-temperature coherence dynamics can be reconstructed exactly from either single pure branch by applying the temperature-dependent contraction factor \(q^n\) and the corresponding phase correction.

We have assumed \(q\neq0\) when defining \(\Theta\).
If \(q=0\), the coherence vanishes after one collision and the phase correction becomes irrelevant.

Finally, we note that the structures of Eqs.~\eqref{eq:zn_thermal} and \eqref{eq:w_th_n} are consistent with the finite-temperature spin-relaxation solution reported in Ref.~\cite{rau1963}. The model considered there is based on an isotropic Heisenberg interaction and therefore constitutes a special case of the present \(U(1)\)-symmetric framework. In both descriptions, the longitudinal component relaxes toward the bath polarization, whereas the transverse components exhibit a temperature-dependent damping accompanied by phase rotation.

\subsection{Implementation workflow on gate-based quantum computers}
\label{subsec:implementation}

\begin{figure}[h]
\centering
\includegraphics[width=1\linewidth]{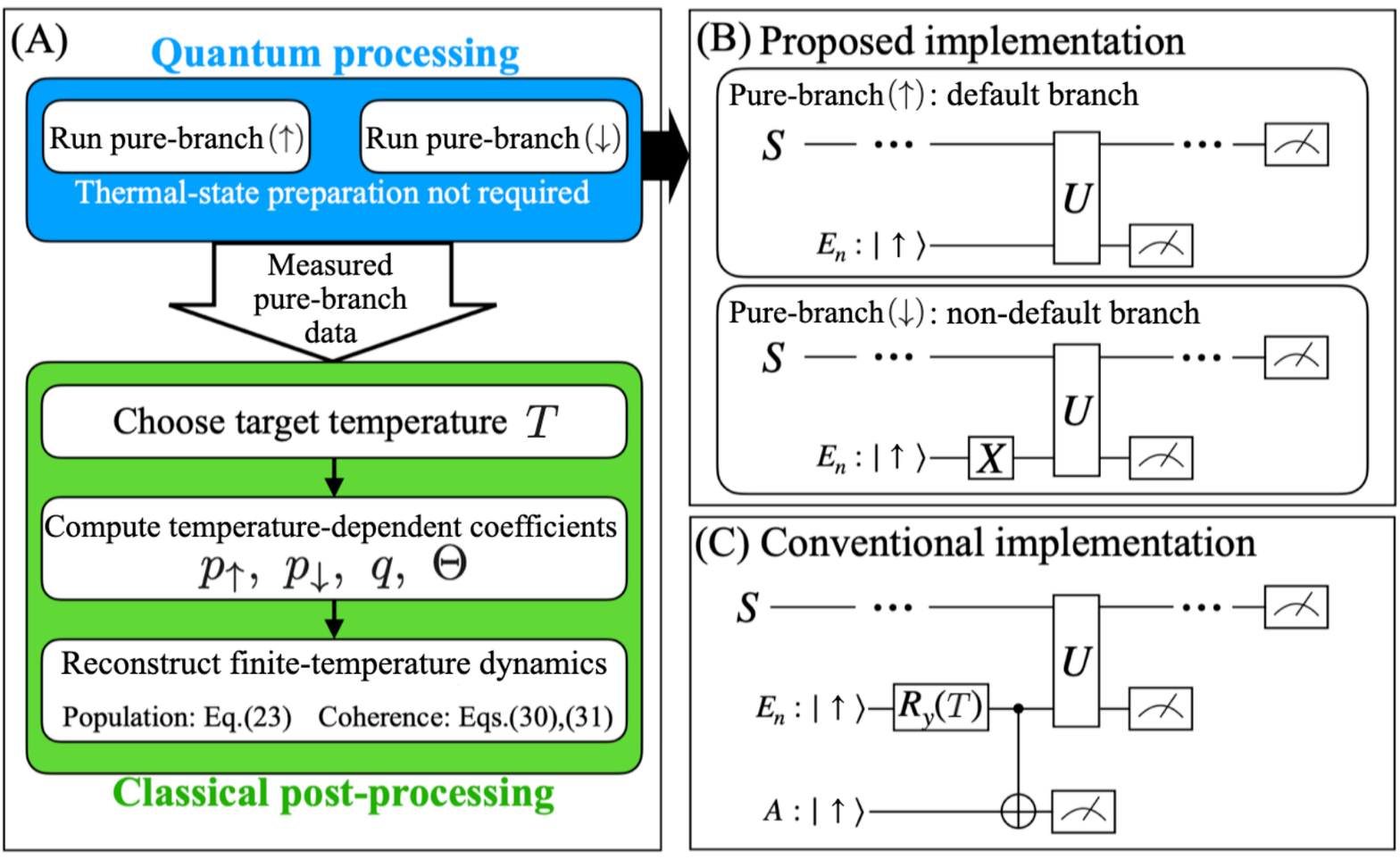}
\vspace{-10mm}
\caption{
(A) Workflow for reconstructing finite-temperature collision dynamics by post-processing pure-branch data.
While only the two pure branches are implemented on the quantum computer, the finite-temperature dynamics at a target temperature \(T\) is reconstructed by classical post-processing:
populations are reconstructed by the convex mixture in Eq.~\eqref{eq:population_exact_reconstruction}, while the coherence dynamics is reconstructed using the contraction factor and phase correction in Eqs.~\eqref{eq:coh_recon_up} and \eqref{eq:coh_recon_down}.
(B) Quantum circuits for implementing the pure branches.
For the hardware convention illustrated here, the default state is taken to be \(|d\rangle=|\!\!\uparrow\rangle\), so that the non-default state \(|\bar d\rangle=|\!\!\downarrow\rangle\) is prepared by applying a single \(X\) gate.
(C) Quantum circuit for a conventional purification-based implementation of the finite-temperature collision model.
In this case, the minimal ancilla-preparation circuit requires an auxiliary qubit, a temperature-dependent single-qubit rotation, and an entangling CNOT gate.
}
\label{fig:workflow}
\end{figure}

Our goal is to reconstruct the finite-temperature collision dynamics
generated by repeated application of the thermal collision map
\(\Phi_{\rm th}\) in Eq.~\eqref{eq:cm_thermal} at a target temperature
\(T\), without preparing the corresponding Gibbs-state ancillas on the
quantum computer.
The target temperature determines the thermal probabilities
\(p_\uparrow\) and \(p_\downarrow\), which enter the reconstruction
formulae derived in Secs.~\ref{subsec:post_population} and \ref{subsec:post_coherence}.
We now translate these formulae into a gate-based implementation
protocol in which the quantum computer is used only to generate the
pure-branch data, while the specification of the target temperature is
deferred to classical post-processing.
The resulting workflow is illustrated in Fig.~\ref{fig:workflow}.

In the collision model considered here, each environmental ancilla \(E_n\) is a single qubit that interacts with the system during the \(n\)th collision.
Before each collision, the ancilla is initialized in one of the two basis states \(|\!\uparrow\rangle\) or \(|\!\downarrow\rangle\), and after the collision it is discarded or reset before being reused for the next collision.
No thermal mixed state needs to be prepared on the quantum computer.

To describe the implementation of these two pure branches independently of a particular hardware convention, we denote by \(|d\rangle\) the basis state obtained directly by hardware initialization or reset and refer to it as the \emph{default} state.
The other basis state is denoted by \(|{\bar d}\rangle\) and referred to as the \emph{non-default} state.
We likewise refer to the corresponding pure branches as the default and non-default branches, respectively.
The default branch can be implemented without any additional ancilla-preparation gate, whereas the non-default branch is obtained by applying a single \(X\) gate to the default state.
The correspondence between \(\{|d\rangle,|{\bar d}\rangle\}\) and the qubit basis states  \(\{|\!\uparrow\rangle,|\!\downarrow\rangle\}\) depends on the hardware convention.
In Fig.~\ref{fig:workflow}, we adopt the convention \(|d\rangle=|\!\uparrow\rangle\) and \(|{\bar d}\rangle=|\!\downarrow\rangle\).

The two reconstruction formulae derived in Secs.~\ref{subsec:post_population} and \ref{subsec:post_coherence} have different practical implications.
For the population sector, Eq.~\eqref{eq:population_exact_reconstruction} requires the pure-branch data from both branches.
Thus, the quantum-processing stage consists of running the collision circuits separately for the default and non-default branches, corresponding to the ancilla states \(|d\rangle\) and \(|{\bar d}\rangle\).
Once these data have been obtained, the finite-temperature population dynamics at any target temperature is constructed classically by weighting the two pure-branch data sets with the corresponding thermal probabilities \(p_\uparrow\) and \(p_\downarrow\).

The coherence sector requires even less pure-branch data.
According to Eqs.~\eqref{eq:coh_recon_up} and \eqref{eq:coh_recon_down}, the coherence data from either pure branch are sufficient.
The coherence data of the chosen branch provide the input for the classical reconstruction, while the factor \(q^n\) and the phase correction \(e^{in(\Theta-\theta_s)}\) are applied classically.
In practice, choosing the default branch for the coherence reconstruction is advantageous because no additional ancilla-preparation gate is then required.

Figure~\ref{fig:workflow}(A) summarizes the complete workflow.
The pure-branch data are first generated on the quantum computer, without preparing a thermal ancilla.
A target temperature \(T\) is then selected in the classical post-processing stage, from which the temperature-dependent quantities \(p_\uparrow\), \(p_\downarrow\), \(q\), and \(\Theta\) are evaluated.
The population and coherence dynamics are subsequently reconstructed using Eq.~\eqref{eq:population_exact_reconstruction} and
Eqs.~\eqref{eq:coh_recon_up}--\eqref{eq:coh_recon_down}, respectively.
Because the pure-branch data contain no dependence on the target temperature, the same data can be reused for different temperatures without rerunning the quantum circuits.

\section{Ancilla-preparation overhead and error analysis}
\label{sec:errorsec}
In this section, we compare post-processing reconstruction with purification as a representative conventional approach to finite-temperature ancilla preparation on gate-based quantum computers.
Purification provides a natural benchmark because it directly realizes a mixed thermal ancilla through a pure-state quantum circuit.
We first compare the preparation overhead of the two approaches and then quantify the resulting preparation-induced errors within the broader error structure of repeated-interaction simulations.

\subsection{Purification-based implementation versus post-processing reconstruction}
\label{subsec:purification_vs_postprocessing}

Section~\ref{subsec:implementation} described how the proposed reconstruction can be implemented using only pure ancillas initialized in the two basis states.
We now compare this implementation with the conventional purification-based preparation of thermal ancillas.

In a purification-based approach, the thermal ancilla state
\begin{equation}
\rho_A^{\rm th}
=
p_\uparrow|\uparrow\rangle\langle\uparrow|
+
p_\downarrow|\downarrow\rangle\langle\downarrow|
\end{equation}
is realized as the reduced state of the purified two-qubit state
\begin{equation}
|\Psi_{\rm th}\rangle
=
\sqrt{p_\uparrow}|\uparrow\uparrow\rangle
+
\sqrt{p_\downarrow}|\downarrow\downarrow\rangle
\label{eq:purified_state}
\end{equation}
by tracing out the auxiliary qubit from the pure state.
A minimal preparation circuit requires a temperature-dependent single-qubit rotation followed by an entangling gate, as illustrated in Fig.~\ref{fig:workflow}(C).
Consequently, the thermal ancilla must be re-prepared before each collision, introducing additional single-qubit and two-qubit gate operations throughout the simulation.

By contrast, the post-processing reconstruction described in Sec.~\ref{subsec:implementation} does not prepare \(\rho_A^{\rm th}\) on the quantum computer.
For the population sector, the two pure branches are generated separately and combined with the thermal probabilities \(p_\uparrow\) and \(p_\downarrow\) in classical post-processing.
For the coherence sector, either one of the pure branches is sufficient, and the finite-temperature dynamics is obtained by applying the classical contraction factor and phase correction derived in Sec.~\ref{subsec:post_coherence}.
In particular, choosing the default branch for the coherence reconstruction requires no additional ancilla-preparation gate.

The distinction between the two implementations is therefore directly reflected in their preparation overhead.
Purification requires a temperature-dependent rotation and an entangling gate for each ancilla preparation.
Post-processing reconstruction, by contrast, requires at most a single \(X\) gate to prepare the non-default pure branch; the default branch requires no additional preparation gate.
This difference is particularly relevant to the coherence reconstruction, for which the default branch alone can be used.
We next quantify how these differences in preparation overhead propagate into the simulated dynamics.

\subsection{Preparation-induced error analysis}
\label{sec:error}

In Ref.~\cite{pocrnic2025}, Pocrnic et al. separated the simulation error in
repeated-interaction simulations into two contributions: the error
associated with implementing the collision dynamics on a quantum
computer and the error arising from approximating the target Lindblad
dynamics by a repeated-interaction collision model. Their analysis
assumes ideal preparation of the thermal ancillas and therefore does
not separately account for ancilla-preparation errors. Here, we extend
this error decomposition by explicitly including imperfect ancilla
preparation as an additional error source.

For convenience, we represent the system state after the $n$th
collision by its Bloch vector $\mathbf{r}_n=(x_n,y_n,z_n)$. We
distinguish four levels of dynamics: the target Lindblad dynamics at
$t=n\tau$, denoted by $\mathbf{r}_n^{\mathrm{L}}$; the ideal
collision-model (CM) dynamics $\mathbf{r}_n^{\mathrm{CM}}$; the
collision-model dynamics including collision-gate (CG) implementation
errors but assuming ideal ancilla preparation,
$\mathbf{r}_n^{\mathrm{CG}}$; and the full quantum-circuit
implementation including both collision-gate and ancilla-preparation
errors, $\mathbf{r}_n^{\mathrm{Full}}$. The total implementation error
can then be decomposed as
\begin{equation}
\left\|
\mathbf{r}_n^{\mathrm{Full}}
-\mathbf{r}_n^{\mathrm{L}}
\right\|
\leq
\left\|
\mathbf{r}_n^{\mathrm{Full}}
-\mathbf{r}_n^{\mathrm{CG}}
\right\|
+
\left\|
\mathbf{r}_n^{\mathrm{CG}}
-\mathbf{r}_n^{\mathrm{CM}}
\right\|
+
\left\|
\mathbf{r}_n^{\mathrm{CM}}
-\mathbf{r}_n^{\mathrm{L}}
\right\|.
\label{eq:error_decomposition}
\end{equation}
The second and third terms correspond, respectively, to the
collision-gate implementation error and the
collision-model-to-Lindblad approximation error considered in
Ref.~\cite{pocrnic2025}. The first term is the additional contribution
considered here and represents the error arising specifically from
imperfect ancilla preparation. Our purpose is not to reevaluate the
other two contributions, but to compare this preparation-induced
contribution between purification-based implementation and
post-processing reconstruction.

The relevant difference between the two implementations is the
ancilla-preparation circuit described in
Sec.~\ref{subsec:purification_vs_postprocessing}. In the
purification-based implementation, each thermal ancilla is prepared
using a temperature-dependent single-qubit rotation followed by an
entangling CNOT gate. We denote the corresponding effective error
probabilities by $p_{\mathrm{R}}$ and $p_{\mathrm{CX}}$, respectively. In
post-processing reconstruction, no thermal mixed state is prepared.
The default pure branch requires no additional ancilla-preparation
gate, whereas the non-default branch requires only a single X gate,
whose effective error probability is denoted by $p_{\mathrm{X}}$. Thus, the
comparison below isolates how this difference in preparation-gate
overhead propagates into the simulated dynamics.

We assume that the effective collision-gate errors and
ancilla-preparation errors are stationary over successive collision
steps. The same collision-gate errors are included in both
$\mathbf{r}_n^{\mathrm{Full}}$ and $\mathbf{r}_n^{\mathrm{CG}}$, so
that only the additional contribution associated with imperfect
ancilla preparation is isolated. We further assume weak preparation
errors that preserve the diagonal form of the ancilla state in the
energy basis to leading order, and retain only the leading
contribution in the preparation error.
A detailed formulation of these assumptions is provided in Appendix~\ref{app:prep_model}.

\begin{table}[t]
\centering
\caption{
Leading-order scaling of preparation-induced errors for
purification-based implementation and post-processing reconstruction
derived in Appendix~\ref{app:scaling}.
Here \(p_{\rm R}\), \(p_{\rm CX}\), and \(p_{\rm X}\) denote the
effective error probabilities associated with the
temperature-dependent rotation, CNOT, and \(X\) gates, respectively.
The scaling orders apply at fixed thermal probabilities and away from
special parameter values at which noise-model-dependent prefactors
vanish.
}
\label{tab:error_summary}
\begin{tabular}{lll}
\hline
Sector & Method & Scaling \\
\hline
Population
& Purification
& \(O(p_{\rm R}+p_{\rm CX})\)
\\
Population
& Post-processing
& \(O(p_{\rm X})\)
\\
Coherence
& Purification
& \(O[n(p_{\rm R}+p_{\rm CX})]\)
\\
Coherence
& Post-processing (\(\bar d\))
& \(O(np_{\rm X})\)
\\
Coherence
& Post-processing (\(d\))
& \(0\)
\\
\hline
\end{tabular}
\end{table}

Under these assumptions, we quantify the preparation-induced
population error by the absolute deviation of the longitudinal Bloch
component,
$|\Delta z_n^{\mathrm{prep}}|
:=|z_n^{\mathrm{Full}}-z_n^{\mathrm{CG}}|$
(see Appendix~\ref{app:population_error}).
For the coherence sector, since the coherence itself generally decays
under repeated collisions, making an absolute error less informative
at large $n$, we instead consider the relative error
$E_n^{\mathrm{prep}}
:=|w_n^{\mathrm{Full}}-w_n^{\mathrm{CG}}|/
|w_n^{\mathrm{CG}}|$ for $w_n^{\mathrm{CG}}\neq0$, where
$w_n^{\mathrm{Full}(\mathrm{CG})}
=x_n^{\mathrm{Full}(\mathrm{CG})}
+i y_n^{\mathrm{Full}(\mathrm{CG})}$
(see Appendix~\ref{app:coherence_error}).
The resulting leading-order scalings are summarized in
Table~\ref{tab:error_summary}.
A distinctive advantage arises in the coherence sector: because
coherence can be reconstructed from either pure branch, choosing the
default branch requires no additional ancilla-preparation gate and
eliminates the preparation-induced contribution considered here.
These scalings are common to the representative preparation-noise
models considered below, whereas their numerical prefactors generally
depend on the preparation-noise mechanism, temperature, and dynamical
parameters (see Eqs.~\eqref{eq:pop_error_pur_dep}, \eqref{eq:coh_error_pur_dep}, \eqref{eq:pop_error_post_dep} and \eqref{eq:coh_error_post_dep} for the depolarizing channel model and Eqs.~\eqref{eq:pop_error_pur_AD}, \eqref{eq:coh_error_pur_AD}, \eqref{eq:pop_error_post_AD} and \eqref{eq:coh_error_post_AD} for the amplitude damping model).

For a quantitative comparison of the two implementations, we next
examine the prefactors underlying the leading-order scalings in
Table~\ref{tab:error_summary}. In the population reconstruction, the
preparation error of the non-default branch is multiplied by the
thermal probability of that branch
(Eq.~\eqref{eq:pop_error_post_general} of
Appendix~\ref{app:population_error_post}), providing an additional
suppression of the population error. In the coherence reconstruction,
by contrast, a single pure branch is used rather than a thermal convex
mixture, so that no such thermal factor appears
(Eq.~\eqref{eq:coh_error_post_general} of
Appendix~\ref{app:coherence_error_post}).

The temperature dependence of these prefactors is also sensitive to
the type of ancilla-preparation noise. We consider two representative models in
Appendix~\ref{app:noise_models}. Depolarizing noise discussed in Sec.~\ref{app:depolarizing} drives the prepared
ancilla toward equal populations of its two energy levels,
corresponding to the high-temperature limit, whereas amplitude damping discussed in Sec.~\ref{app:amplitude_damping}
drives it toward the ground state, corresponding to the
low-temperature limit. The preparation-induced error is therefore
suppressed near the respective temperature limits, while away from
these limits the two models show qualitatively similar behavior,
differing in their numerical prefactors.

\begin{figure}[t]
\centering
\includegraphics[width=1\linewidth]{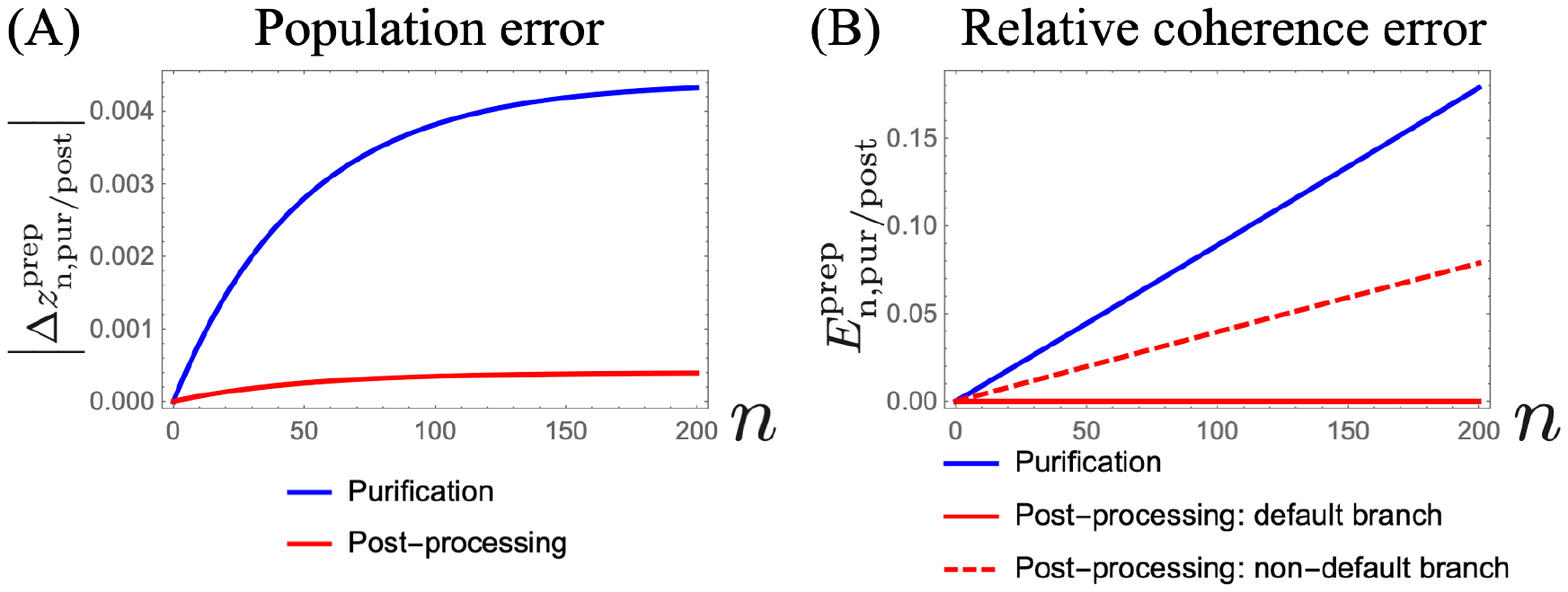}
\vspace{-70pt}
\caption{
Preparation-induced errors as a function of the collision step \(n\)
for the representative amplitude-damping model.
(A) Magnitude of the population preparation error.
The purification-based and post-processing errors both saturate with
increasing \(n\), while post-processing reconstruction reduces the
additional preparation-gate contribution from
\(O(p_{\rm R}+p_{\rm CX})\) to \(O(p_{\rm X})\).
(B) Relative coherence preparation error.
For purification-based implementation, the error grows approximately
as \(O[n(p_{\rm R}+p_{\rm CX})]\), whereas post-processing
reconstruction using the non-default branch gives \(O(np_{\rm X})\).
Using the default branch introduces no additional preparation-gate
error and therefore gives \(E_{n,\rm post,d}^{\rm prep}=0\).
The parameters are
\(p_{\rm R}=10^{-3}\),
\(p_{\rm CX}=10^{-2}\),
\(p_{\rm X}=10^{-3}\),
\(p_\uparrow=0.8\),
\(p_\downarrow=0.2\), and
\(\widetilde{\gamma}=0.98\).
The hardware convention is
\(|d\rangle=|\!\uparrow\rangle\) and
\(|\bar d\rangle=|\!\downarrow\rangle\), with
\(|\!\uparrow\rangle\) identified as the ancilla ground state.
}
\label{fig:prep_error}
\end{figure}

As a representative example, we show the results for amplitude
damping at an intermediate temperature in
Fig.~\ref{fig:prep_error}, where the behavior is not dominated by the
limiting suppression discussed above. The curves are evaluated from
the expressions derived in Appendix~\ref{app:amplitude_damping}
(Eqs.~\eqref{eq:pop_error_pur_AD},
\eqref{eq:coh_error_pur_AD},
\eqref{eq:pop_error_post_AD},
\eqref{eq:coh_error_post_AD}, and
\eqref{eq:coh_error_default_AD}).
We choose $p_\uparrow=0.8$ and $p_\downarrow=0.2$, corresponding to
$z_A=0.6$. The effective gate-error probabilities are set to
$p_R=p_X=10^{-3}$ and $p_{\mathrm{CX}}=10^{-2}$ as illustrative
values, with their orders of magnitude motivated by the custom
noise-model example of Qiskit Aer~\cite{QiskitAerNoise}.
We further set the effective population-relaxation coefficient to
$\widetilde{\gamma}=0.98$, representing weak, contractive relaxation
per collision.

Panel (A) shows the preparation-induced population error.
For both implementations, the error increases with the number of
collisions and eventually saturates. For the parameters used here,
post-processing reconstruction reduces the saturated error by
approximately one order of magnitude.
This quantitative reduction reflects both the reduced
preparation-gate overhead and the thermal-probability-dependent
prefactors discussed above.

Panel (B) shows the relative preparation-induced coherence error.
For purification and reconstruction using the non-default branch, the
error increases approximately linearly with the number of collisions,
as expected from Table~\ref{tab:error_summary}. Unlike in the
population sector, the non-default-branch contribution is not
multiplied by its thermal probability, and the purification and
non-default-branch curves are consequently closer to each other than
the corresponding population-error curves in panel (A).
The default-branch result remains zero because no additional
ancilla-preparation gate is required.

We also examined the temperature dependence within the intermediate
temperature regime, although the corresponding numerical results are
not shown here. The observed trends are consistent with the analytical
expressions in Appendix~\ref{app:amplitude_damping}.
In the population sector, the preparation-induced
errors increase with temperature for both purification and
post-processing reconstruction, while their relative separation
changes only weakly with temperature (see Eqs.~\eqref{eq:pop_error_pur_AD} and \eqref{eq:pop_error_post_AD}).
In the coherence sector, the error increases more strongly with temperature for purification,
whereas post-processing reconstruction continues to suppress the
non-default-branch contribution (see Eqs.~\eqref{eq:coh_error_pur_AD} and \eqref{eq:coh_error_post_AD}).
The default-branch coherence contribution remains zero independently of temperature (see Eq.~\eqref{eq:coh_error_default_AD}).

It is also useful to place the preparation-error reduction discussed
above in the broader context of the collision-model discretization
considered by Pocrnic et al.~\cite{pocrnic2025}. At fixed total
evolution time, refining the discretization reduces the
collision-model-to-Lindblad correspondence error by shortening each
collision interval. On noisy quantum hardware, however, it also
increases the number of collision circuits and hence the exposure to
physical gate errors. These two effects therefore exhibit competing
trends. Because ancilla-preparation operations are repeated at every
collision, post-processing reconstruction mitigates one source of
hardware error that becomes increasingly relevant as the
discretization is refined. In this sense, the present approach
complements the systematic convergence toward Lindblad dynamics
established for repeated-interaction simulations.

\section{Discussion}
\label{sec:discussion}

The applicability of post-processing reconstruction is closely related to the dynamical structure imposed by the system--ancilla interaction.
In the qubit--qubit setting considered here, the $U(1)$ symmetry, $[H,\sigma_z^S+\sigma_z^A]=0$, corresponds physically to conservation of the total excitation number during each collision.
The $U(1)$-symmetric framework developed here therefore naturally describes excitation-conserving interactions, including exchange-type couplings commonly used in collision models of thermal relaxation.
Interactions that do not conserve the total excitation number generally fall outside this symmetry class.
As discussed below, however, the role of this symmetry differs between the coherence and population sectors.

At the level of the reduced dynamics, the $U(1)$ symmetry constrains each pure branch to act on the transverse Bloch vector through a single complex multiplier, $w\mapsto\lambda_s w$, and, crucially, enforces a common contraction factor for the two branches while allowing branch-dependent phases.
This structure enables the finite-temperature coherence to be reconstructed from either single pure branch by applying a classical contraction factor and phase correction.
Without this symmetry, the transverse dynamics may generally mix $w$ and $w^*$, and neither the common-contraction structure nor the single-branch
coherence reconstruction is guaranteed.

The population reconstruction is less restrictive and does not require the full $U(1)$ symmetry itself.
A sufficient condition for the convex-mixture reconstruction is that the population forms a closed affine sector and that the two pure-branch population maps have the same linear part,
\begin{equation}
 \Phi_s: z \mapsto \gamma z + c_s,
 \qquad s=\uparrow,\downarrow,
\end{equation}
with a branch-independent coefficient $\gamma$.
Under this condition, the branch dependence enters only through the affine shift $c_s$, and the finite-temperature population satisfies
\begin{equation}
 z_n^{\rm th}
 =
 p_\uparrow z_n^{(\uparrow)}
 +
 p_\downarrow z_n^{(\downarrow)}
\end{equation}
for arbitrary collision number $n$.
Physically, this means that changing the ancilla state may change the population toward which the system relaxes, but not the relaxation factor itself.
This condition can hold even in the absence of the full $U(1)$ symmetry, so the population reconstruction can in principle apply to a broader class of collision models than the coherence reconstruction.

The present analysis also isolates only the error arising from imperfect ancilla preparation.
Other experimental imperfections, including collision-gate errors, reset errors, readout errors, finite-shot sampling noise, and uncertainties in the reconstructed parameters, remain in practical implementations and should be analyzed separately.

\section{Conclusion}
\label{sec:conclusion}

We have introduced post-processing reconstruction, a framework for simulating finite-temperature collision dynamics from temperature-independent pure-branch data obtained with pure ancillas.
Within a broad class of \(U(1)\)-symmetric system--ancilla interactions, the quantum-processing stage becomes independent of the target temperature, which enters only through classical post-processing.

This reconstruction naturally separates into population and coherence sectors.
The population dynamics is recovered exactly as a convex mixture of the two pure-branch dynamics, whereas the coherence dynamics is reconstructed from the dynamics of a single pure branch by applying an analytically determined contraction factor and phase correction.
Together, these reconstruction rules enable the finite-temperature collision dynamics to be recovered without explicitly preparing thermal ancilla states.

From the viewpoint of quantum implementation, post-processing reconstruction provides two practical advantages.
First, the same pure-branch data can be reused to reconstruct dynamics at different target temperatures, enabling temperature retargeting without rerunning the quantum circuits.
Second, by eliminating explicit thermal-state preparation, the method removes the associated ancilla-preparation overhead and reduces the corresponding preparation-induced errors compared with purification-based implementations.

The present results demonstrate that, for an important class of collision models, finite-temperature open-system dynamics can be simulated by separating temperature-independent quantum processing from temperature-dependent classical reconstruction.
These results motivate further exploration of whether analogous reconstruction strategies can be developed by exploiting dynamical structures associated with other symmetries.
Another important direction is the extension to non-Markovian collision models, where memory retained across successive collisions modifies the dynamical structure underlying the present reconstruction.
It remains an open question to determine which elements of the present reconstruction strategy can be retained in such memory-bearing settings and what additional dynamical conditions are required.

\section*{Declarations}

\subsection*{Funding}
This work was supported by JSPS KAKENHI Grant Nos.~22K13971 and 22K03467.

\subsection*{Author contributions}
Kazunari Hashimoto and Chikako Uchiyama contributed equally to the
conception and development of this work. Both authors contributed to
the theoretical analysis, interpretation of the results, and preparation
of the manuscript. Both authors read and approved the final manuscript.

\subsection*{Competing interests}
The authors declare no competing interests.

\subsection*{Data availability}
The data supporting the findings of this study are available from the
corresponding author upon reasonable request.

\subsection*{Code availability}
The code used to generate the numerical results presented in this study
is available from the corresponding author upon reasonable request.

\appendix

\section{Derivation of the sector decomposition}
\label{app:covariance}

This appendix provides the derivation of the sector decomposition used in Sec.~\ref{sec:model}.
We show that the \(U(1)\) symmetry of the system--ancilla interaction implies that the reduced collision map has a block-diagonal affine form in the Bloch representation, leading to the separation of the population and coherence sectors.

Let
\begin{equation}
U_\phi^S
:=
e^{-i\phi\sigma_z^S/2},
\qquad
U_\phi^A
:=
e^{-i\phi\sigma_z^A/2},
\end{equation}
and suppose that the system--ancilla Hamiltonian satisfies
\begin{equation}
[H,\sigma_z^S+\sigma_z^A]=0.
\label{eq:u1sym_app}
\end{equation}
Since the collision unitary is
\(U=e^{-iH\tau}\),
Eq.~\eqref{eq:u1sym_app} immediately implies
\begin{equation}
[U,U_\phi^S\otimes U_\phi^A]=0.
\end{equation}

We further assume that the ancilla state is diagonal in the
\(\sigma_z\) basis,
\begin{equation}
[\rho_A,U_\phi^A]=0.
\end{equation}
Using these relations together with the cyclic property of the partial trace, the reduced collision map satisfies
\begin{equation}
\Phi
\!\left[
U_\phi^S
\rho_S
U_\phi^{S\dagger}
\right]
=
U_\phi^S
\Phi[\rho_S]
U_\phi^{S\dagger},
\end{equation}
which establishes the \(U(1)\)-covariance of the collision dynamics.

In the Bloch representation,
\[
\mathbf r\mapsto T\mathbf r+\mathbf t,
\]
the \(U(1)\)-covariance requires
\begin{equation}
TR_z(\phi)
=
R_z(\phi)T,
\qquad
\mathbf t
=
R_z(\phi)\mathbf t
\end{equation}
for any rotation angle \(\phi\), where the rotation matrix is defined by
\begin{equation}
R_z(\phi):=
\begin{pmatrix}
\cos\phi & -\sin\phi & 0\\
\sin\phi & \cos\phi & 0\\
0&0&1
\end{pmatrix}.
\end{equation}
The only affine map satisfying these relations is
\begin{equation}
T
=
\begin{pmatrix}
\mu & \nu & 0\\
-\nu & \mu & 0\\
0&0&\gamma
\end{pmatrix},
\qquad
\mathbf t
=
\begin{pmatrix}
0\\
0\\
f
\end{pmatrix}.
\label{eq:block_affine}
\end{equation}

Equation~\eqref{eq:block_affine} shows that the longitudinal (\(z\)) component evolves independently of the transverse (\(x,y\)) components.
This block-diagonal structure is the origin of the population and coherence sector decomposition used throughout the main text.

\section{Derivation of the pure-branch collision maps}
\label{app:pure_branch_maps}

This appendix derives the explicit collision maps for the two pure ancilla branches introduced in Sec.~\ref{subsec:pure_branch_maps}.
These maps constitute the building blocks of the post-processing reconstruction developed in the main text.

\subsection{Population branch maps}

We employ the ordered basis
\[
|\uparrow\uparrow\rangle,\;
|\uparrow\downarrow\rangle,\;
|\downarrow\uparrow\rangle,\;
|\downarrow\downarrow\rangle.
\]
The \(U(1)\) symmetry implies conservation of the total excitation number, so that the Hilbert space decomposes as
\[
\mathcal H
=
\mathcal H_2
\oplus
\mathcal H_1
\oplus
\mathcal H_0,
\]
where
\[
\mathcal H_2
=
{\rm span}\{|\uparrow\uparrow\rangle\},
\]
\[
\mathcal H_1
=
{\rm span}
\{
|\uparrow\downarrow\rangle,
|\downarrow\uparrow\rangle
\},
\]
and
\[
\mathcal H_0
=
{\rm span}
\{
|\downarrow\downarrow\rangle
\}.
\]
The most general unitary compatible with this decomposition is
\begin{equation}
U:=
\begin{pmatrix}
e^{-i\phi_{\uparrow}} & 0 & 0 & 0\\
0 & e^{i(\chi+\alpha)}\cos\delta
  & e^{i(\chi+\beta)}\sin\delta & 0\\
0 & -e^{i(\chi-\beta)}\sin\delta
  & e^{i(\chi-\alpha)}\cos\delta & 0\\
0 & 0 & 0 & e^{-i\phi_{\downarrow}}
\end{pmatrix}.
\label{eq:general_u1_unitary}
\end{equation}

For a pure ancilla branch \(s=\uparrow,\downarrow\), the reduced collision map is
\[
\Phi_s[\rho_S]
:=
{\rm Tr}_A
\!\left[
U
(\rho_S\otimes|s\rangle\langle s|)
U^\dagger
\right].
\]
Straightforward evaluation gives
\begin{align}
\Phi_\uparrow:
\qquad
z
&\longmapsto
\cos^2\delta\,z+\sin^2\delta,
\\
\Phi_\downarrow:
\qquad
z
&\longmapsto
\cos^2\delta\,z-\sin^2\delta.
\end{align}
Introducing
\[
\gamma
:=
\cos^2\delta,
\]
the two branch maps become
\begin{align}
\Phi_\uparrow:
\qquad
z
&\longmapsto
\gamma z+(1-\gamma),
\\
\Phi_\downarrow:
\qquad
z
&\longmapsto
\gamma z-(1-\gamma).
\end{align}
The two branches therefore possess the same linear relaxation coefficient while differing only in the sign of the affine shift.
This immediately leads to the convex-mixture reconstruction of the population dynamics derived in Sec.~\ref{sec:postprocessing}.

\subsection{Coherence branch maps}

We next consider the transverse coherence
\(
w=x+iy,
\)
which is represented by the system raising operator
\(
\sigma_+^S
:=
|\uparrow\rangle
\langle\downarrow|.
\)
Evaluating the reduced collision map gives
\begin{align}
\Phi_\uparrow[\sigma_+^S]
&=
e^{-i(\phi_\uparrow+\chi-\alpha)}
\cos\delta\,
\sigma_+^S,
\\
\Phi_\downarrow[\sigma_+^S]
&=
e^{i(\phi_\downarrow+\chi+\beta)}
\cos\delta\,
\sigma_+^S.
\end{align}
Hence,
\begin{align}
\lambda_\uparrow
&:=
\cos\delta\,
e^{-i(\phi_\uparrow+\chi-\alpha)},
\\
\lambda_\downarrow
&:=
\cos\delta\,
e^{i(\phi_\downarrow+\chi+\beta)}.
\end{align}
Using \(\gamma=\cos^2\delta\), the branch multipliers may be written
without loss of generality as
\begin{equation}
\lambda_s=\sqrt{\gamma}\,e^{i\theta_s},
\qquad
s=\uparrow,\downarrow,
\end{equation}
where
\begin{align}
\theta_\uparrow
&:=
-(\phi_\uparrow+\chi-\alpha),
\\
\theta_\downarrow
&:=
\phi_\downarrow+\chi+\beta.
\end{align}
Here, the sign of \(\cos\delta\) is absorbed into the phase
\(\theta_s\).

The coherence therefore evolves according to
\[
w
\longmapsto
\lambda_s w.
\]
Equivalently, the transverse Bloch matrix is
\[
T_s
=
R_z(\theta_s)
\Gamma,
\]
where
\[
\Gamma
:=
{\rm diag}
(
\sqrt{\gamma},
\sqrt{\gamma},
\gamma
).
\]
The two pure branches therefore share the same dissipative contraction while differing only in their phase rotations.
These structural properties constitute the basis of the coherence reconstruction formula derived in Sec.~\ref{sec:postprocessing}.

\section{Ancilla-preparation errors}
\label{app:prep_model}

In this Appendix, we derive the preparation-induced error expressions
summarized in Sec.~\ref{sec:error}.
Our purpose is to isolate the preparation-induced contribution from the
errors generated by the collision-gate implementation itself.
We first derive, without specifying a particular noise model, how a small
preparation-induced deviation of the ancilla polarization propagates through
the repeated collision dynamics.
The resulting expressions are obtained separately for the population and
coherence sectors.
We then evaluate these general expressions for two representative
preparation-noise models, depolarizing noise and amplitude damping, thereby
illustrating how the prefactors and their temperature dependence depend on
the underlying noise mechanism.
Throughout this Appendix, we make the following assumptions.

\begin{enumerate}
\item
We isolate the ancilla-preparation contribution using the error decomposition
introduced in Sec.~\ref{sec:error}.
We denote by $\mathbf{r}_n^{\rm Full}$ the dynamics in which both the
collision-gate implementation and ancilla preparation are imperfect, and by
$\mathbf{r}_n^{\rm CG}$ the dynamics generated by the same collision-gate
implementation but with ideal ancilla preparation.
The latter is taken as the reference dynamics in this Appendix.
Accordingly, the preparation-induced error is defined by
\begin{equation}
    \Delta \mathbf{r}_n^{\rm prep}
    :=
    \mathbf{r}_n^{\rm Full}
    -
    \mathbf{r}_n^{\rm CG}.
    \label{eq:prep_error_definition}
\end{equation}
This quantity corresponds to the first term on the right-hand side of
Eq.~\eqref{eq:error_decomposition}.

\item
We assume that the effective errors are stationary over successive collision
steps.
The errors of the collision-gate implementation are therefore represented by
collision-step-independent effective parameters
$\widetilde{\gamma}$ and $\widetilde{\lambda}_s$ with
$s=\uparrow,\downarrow$.
Likewise, the errors associated with the ancilla-preparation gates are
represented by collision-step-independent effective error probabilities
$p_G$, with $G={\rm R},{\rm CX},{\rm X}$.

\item
We assume that the effective preparation error can be described by a small
deviation of the longitudinal ancilla polarization
\begin{equation}
    \widetilde{z}
    :=
    z+\delta z,
    \qquad
    |\delta z|\ll 1,
    \label{eq:generic_delta_z}
\end{equation}
and we retain terms only to first order in $\delta z$.
For the purification-based implementation,
\(
    \widetilde{z}_A
    :=
    z_A+\delta z_A,
\)
with
\(
    z_A=p_\uparrow-p_\downarrow,
\)
whereas for a pure branch $s$ used in post-processing reconstruction,
\(
    \widetilde{z}_s
    :=
    z_s+\delta z_s,
\)
with
\(
    z_s=\pm1.
\)
The default branch requires no additional preparation gate, and hence
\(
    \delta z_d=0.
\)
At this stage, no particular relation between $\delta z$ and the effective
error probability $p_G$ is assumed.
\end{enumerate}

\subsection{Population dynamics}
\label{app:population_error}

We first consider the propagation of the ancilla-preparation error in
the population sector.
For the class of collision models considered in the present work, the
population dynamics generated by the noisy collision-gate implementation retains
the affine form
\begin{equation}
    z_{n+1}
    =
    \widetilde{\gamma} z_n
    +
    (1-\widetilde{\gamma})z_{\rm anc},
    \label{eq:population_affine_app}
\end{equation}
where $z_{\rm anc}=z_A$ for ideal ancilla preparation or $z_{\rm anc}={\tilde z}_A$ for imperfect ancilla preparation.
The parameter $\widetilde{\gamma}$ already contains the effect of the
collision-gate implementation errors and is common to the full and reference
dynamics.

\subsubsection{Purification-based implementation}

For ideal ancilla preparation, the reference dynamics is
\begin{equation}
    z_{n+1}^{\rm CG}
    =
    \widetilde{\gamma}z_n^{\rm CG}
    +
    (1-\widetilde{\gamma})z_A .
    \label{eq:pop_ref_pur}
\end{equation}
When the reduced thermal ancilla produced by the purification circuit
has the polarization
$\widetilde z_A=z_A+\delta z_A$, the corresponding full dynamics is
\begin{equation}
    z_{n+1}^{\rm Full}
    =
    \widetilde{\gamma}z_n^{\rm Full}
    +
    (1-\widetilde{\gamma})
    (z_A+\delta z_A).
    \label{eq:pop_full_pur}
\end{equation}
Defining
\begin{equation}
    \Delta z_{n,\rm pur}^{\rm prep}
    =
    z_n^{\rm Full}-z_n^{\rm CG},
\end{equation}
we obtain
\begin{equation}
    \Delta z_{n+1,\rm pur}^{\rm prep}
    =
    \widetilde{\gamma}
    \Delta z_{n,\rm pur}^{\rm prep}
    +
    (1-\widetilde{\gamma})\delta z_A.
    \label{eq:pop_recursion_pur}
\end{equation}
Since the initial system state is common to the two dynamics,
$\Delta z_{0,\rm pur}^{\rm prep}=0$, hence, repeated application of the recurrence equation gives
\begin{equation}
    \Delta z_{n,\rm pur}^{\rm prep}
    =
    (1-\widetilde{\gamma}^{\,n})\delta z_A .
    \label{eq:pop_error_pur_general}
\end{equation}

\subsubsection{Post-processing reconstruction}
\label{app:population_error_post}

In the post-processing reconstruction, the thermal population dynamics is
reconstructed from the two pure branches.
Because the default branch requires no additional preparation
gate, its preparation-induced contribution vanishes,
$\delta z_d=0$.
For the non-default branch, the imperfectly prepared polarization is
\begin{equation}
    \widetilde z_{\bar d}
    =
    z_{\bar d}
    +
    \delta z_{\bar d}.
\end{equation}
The corresponding non-default branch population therefore differs from its
reference value by
\begin{equation}
    \Delta z_{n,\bar d}^{\rm prep}
    =
    (1-\widetilde{\gamma}^{\,n})
    \delta z_{\bar d}.
    \label{eq:pop_branch_error_general}
\end{equation}
Since the thermal population is reconstructed as the convex
combination of the two pure-branch populations, only the non-default
branch contributes to the preparation error.
We therefore obtain
\begin{equation}
    \Delta z_{n,\rm post}^{\rm prep}
    =
    p_{\bar d}
    (1-\widetilde{\gamma}^{\,n})
    \delta z_{\bar d},
    \label{eq:pop_error_post_general}
\end{equation}
where $p_{\bar d}$ is the thermal probability associated with the non-default state $\ket{\bar d}$.

\subsection{Coherence dynamics}
\label{app:coherence_error}

We next consider the coherence sector.
Under the assumptions stated above, the ancilla state immediately
before the collision remains diagonal in the energy basis.
Hence, for both the purification-based implementation and post-processing reconstruction,
\begin{equation}
    x_A=y_A=0.
\end{equation}
The preparation error therefore does not affect the system coherence
through an ancilla transverse component.
Instead, it acts indirectly through the error in the longitudinal
polarization, which changes the relative contribution of the
$|\!\!\uparrow\rangle$ and $|\!\!\downarrow\rangle$ collision branches.

Since a diagonal ancilla with polarization \(z\) can be written as
\begin{equation}
\rho_A(z)=\frac{1+z}{2}|\!\uparrow\rangle\!\langle\uparrow\!|+\frac{1-z}{2}|\!\downarrow\rangle\!\langle\downarrow\!|,
\end{equation}
linearity of the reduced collision map gives the effective coherence multiplier as
\begin{equation}
    \widetilde{\lambda}(z)
    :=
    \frac{1+z}{2}\widetilde{\lambda}_\uparrow
    +
    \frac{1-z}{2}\widetilde{\lambda}_\downarrow.
    \label{eq:lambda_general_z}
\end{equation}
In particular, for the thermal ancilla, 
\begin{equation}
    \widetilde{\lambda}_{\rm th}
    =
    p_\uparrow\widetilde{\lambda}_\uparrow
    +
    p_\downarrow\widetilde{\lambda}_\downarrow
    =
    \widetilde{\lambda}(z_A).
    \label{eq:lambda_th_app}
\end{equation}

Because the magnitude of the coherence generally decays under repeated
collisions, the absolute difference in $w_n=x_n+iy_n$ also tends to
decrease at long times.
We therefore characterize the preparation-induced coherence error by
the relative deviation
\begin{equation}
    E_n^{\rm prep}
    =
    \left|
    \frac{
    w_n^{\rm Full}-w_n^{\rm CG}
    }{
    w_n^{\rm CG}
    }
    \right|,
    \label{eq:relative_coherence_error}
\end{equation}
whenever $w_n^{\rm CG}\neq0$.

\subsubsection{Purification-based implementation}

For the reference dynamics,
\begin{equation}
    w_n^{\rm CG}
    =
    \widetilde{\lambda}_{\rm th}^{n}w_0
    =
    [\widetilde{\lambda}(z_A)]^{\,n}w_0 .
    \label{eq:coh_ref_pur}
\end{equation}
For the full dynamics, the preparation-induced polarization shift
$z_A\rightarrow z_A+\delta z_A$ changes the multiplier by
\begin{align}
    \delta\widetilde{\lambda}_{\rm th}
    &=
    \widetilde{\lambda}(z_A+\delta z_A)
    -
    \widetilde{\lambda}(z_A)
    \nonumber\\
    &=
    \frac{\delta z_A}{2}
    \left(
    \widetilde{\lambda}_\uparrow
    -
    \widetilde{\lambda}_\downarrow
    \right).
    \label{eq:delta_lambda_pur_general}
\end{align}
Expanding the repeated dynamics to first order in $\delta {\tilde\lambda}_{\rm th}$ gives
\begin{align}
    w_n^{\rm Full}
    &=
    \left(
    \widetilde{\lambda}_{\rm th}
    +
    \delta\widetilde{\lambda}_{\rm th}
    \right)^n w_0
    \nonumber\\
    &\simeq
    \left[
    \widetilde{\lambda}_{\rm th}^{\,n}
    +
    n\widetilde{\lambda}_{\rm th}^{\,n-1}
    \delta\widetilde{\lambda}_{\rm th}
    \right]w_0 .
\end{align}
Consequently,
\begin{equation}
    E_{n,\rm pur}^{\rm prep}
    \simeq
    \frac{n}{2}
    \left|
    \delta z_A
    \frac{
    \widetilde{\lambda}_\uparrow
    -
    \widetilde{\lambda}_\downarrow
    }{
    \widetilde{\lambda}_{\rm th}
    }
    \right|.
    \label{eq:coh_error_pur_general}
\end{equation}

\subsubsection{Post-processing reconstruction}
\label{app:coherence_error_post}

For a pure branch $s$, the reference coherence dynamics is
\begin{equation}
    w_n^{(s),{\rm CG}}
    =
    \widetilde{\lambda}_s^{\,n}w_0 .
\end{equation}
The thermal coherence is reconstructed from one of the pure branches
by multiplying the measured branch coherence by the corresponding
classical reconstruction factor.
Since this factor is determined from the ideal collision parameters,
the additional preparation error is entirely inherited from the
deviation of the measured branch dynamics.

For the non-default branch,
$z_{\bar d}\rightarrow z_{\bar d}+\delta z_{\bar d}$ changes the
branch coherence multiplier by
\begin{equation}
    \delta\widetilde{\lambda}_{\bar d}
    =
    \frac{\delta z_{\bar d}}{2}
    \left(
    \widetilde{\lambda}_\uparrow
    -
    \widetilde{\lambda}_\downarrow
    \right).
    \label{eq:delta_lambda_post_general}
\end{equation}
To first order in $\delta z_{\bar d}$ gives
\begin{equation}
    E_{n,\rm post,\bar d}^{\rm prep}
    \simeq
    \frac{n}{2}
    \left|
    \delta z_{\bar d}
    \frac{
    \widetilde{\lambda}_\uparrow
    -
    \widetilde{\lambda}_\downarrow
    }{
    \widetilde{\lambda}_{\bar d}
    }
    \right|.
    \label{eq:coh_error_post_general}
\end{equation}

In contrast to the non-default branch, the default branch requires no additional preparation gate
and therefore gives
\begin{equation}
    E_{n,\rm post,d}^{\rm prep}=0.
    \label{eq:coh_error_default}
\end{equation}

\subsection{Evaluation for specific preparation-noise models}
\label{app:noise_models}

We now relate the polarization deviations $\delta z_A$ and
$\delta z_{\bar d}$ appearing in
Eqs.~\eqref{eq:pop_error_pur_general}--\eqref{eq:coh_error_post_general}
to the effective preparation-gate error probabilities
$p_{\rm R}$, $p_{\rm CX}$, and $p_{\rm X}$.
For this purpose, we consider two representative preparation-noise
models: a depolarizing (DP) channel and an amplitude-damping (AD) channel~\cite{NielsenChuang}.
These models are not intended as microscopic descriptions of a
specific device~\cite{Magesan2013}.
Rather, they provide representative examples of unital and nonunital
noise with different fixed points, allowing us to illustrate how the temperature dependence of the preparation-induced error varies with the underlying noise mechanism~\cite{Wallman2016,Urbanek2021}.
For the purification circuit, $p_{\rm R}$ and $p_{\rm CX}$ denote the
effective error probabilities associated with the temperature-dependent
rotation ($R$) and the CNOT gate, respectively.
For the post-processing reconstruction, $p_{\rm X}$ denotes the
corresponding effective error probability associated with the $X$ gate
used to prepare the non-default branch.

\subsubsection{Depolarizing (DP) channel}
\label{app:depolarizing}

We first consider the effective depolarizing channel
\begin{equation}
    \mathcal{D}_{p_G}(\rho)
    =
    (1-p_G)\rho
    +
    p_G\frac{I}{2},
    \label{eq:depolarizing_app}
\end{equation}
which contracts all Bloch-vector components according to
\begin{equation}
    \mathcal{D}_{p_G}:
    {\mathbf r}
    \rightarrow
    (1-p_G){\mathbf r}.
    \label{eq:dep_bloch}
\end{equation}

Within the effective reduced-ancilla description adopted here,
the successive contractions associated with the R and CNOT gates
give
\begin{equation}
    \widetilde z_A^{\rm DP}
    =
    (1-p_{\rm R})(1-p_{\rm CX})z_A.
\end{equation}
Hence, to first order in the gate-error probabilities,
\begin{equation}
    \delta z_A^{\rm DP}
    :=\widetilde z_A^{\rm DP}-z_A
    \simeq
    -(p_{\rm R}+p_{\rm CX})z_A.
    \label{eq:delta_z_pur_dep}
\end{equation}

For the purification-based implementation, substituting
Eq.~\eqref{eq:delta_z_pur_dep} into the general population-error
formula, Eq.~\eqref{eq:pop_error_pur_general}, gives
\begin{equation}
    \Delta z_{n,\rm pur}^{\rm DP}
    \simeq
    -(1-\widetilde{\gamma}^{\,n})
    (p_{\rm R}+p_{\rm CX})z_A .
    \label{eq:pop_error_pur_dep}
\end{equation}
Likewise, substitution into the general coherence-error formula,
Eq.~\eqref{eq:coh_error_pur_general}, yields
\begin{equation}
    E_{n,\rm pur}^{\rm DP}
    \simeq
    \frac{n}{2}
    (p_{\rm R}+p_{\rm CX})
    |z_A|
    \left|
    \frac{
    \widetilde{\lambda}_\uparrow
    -
    \widetilde{\lambda}_\downarrow
    }{
    \widetilde{\lambda}_{\rm th}
    }
    \right|.
    \label{eq:coh_error_pur_dep}
\end{equation}

For the non-default branch of the post-processing reconstruction,
the $X$-gate error gives
\begin{equation}
    \delta z_{\bar d}^{\rm DP}
    \simeq
    -p_{\rm X}z_{\bar d}.
    \label{eq:delta_z_post_dep}
\end{equation}
Substituting Eq.~\eqref{eq:delta_z_post_dep} into the general
post-processing population-error formula,
Eq.~\eqref{eq:pop_error_post_general}, gives
\begin{equation}
    \Delta z_{n,\rm post}^{\rm DP}
    \simeq
    -(1-\widetilde{\gamma}^{\,n})
    p_{\bar d}p_{\rm X}z_{\bar d}.
    \label{eq:pop_error_post_dep}
\end{equation}
Similarly, substitution into the corresponding general
coherence-error formula, Eq.~\eqref{eq:coh_error_post_general},
gives
\begin{equation}
    E_{n,\rm post,\bar d}^{\rm DP}
    \simeq
    \frac{n}{2}
    p_{\rm X}
    \left|
    z_{\bar d}
    \frac{
    \widetilde{\lambda}_\uparrow
    -
    \widetilde{\lambda}_\downarrow
    }{
    \widetilde{\lambda}_{\bar d}
    }
    \right|.
    \label{eq:coh_error_post_dep}
\end{equation}
Since $|z_{\bar d}|=1$, the polarization-dependent factor in the
last expression is of unit magnitude.

For the default branch, no ancilla-preparation gate is required,
so that $\delta z_d=0$. The general result
Eq.~\eqref{eq:coh_error_default} therefore directly gives
\begin{equation}
    E_{n,\mathrm{post},d}^{\rm DP}=0.
    \label{eq:coh_error_default_DP}
\end{equation}

The temperature dependence of these expressions reflects the fixed
point of the depolarizing channel. Since this channel is unital and
leaves the maximally mixed state invariant, the thermal ancilla
approaches its fixed point as $T\to\infty$, for which $z_A\to0$.
Accordingly, Eqs.~\eqref{eq:pop_error_pur_dep} and
\eqref{eq:coh_error_pur_dep} show that both the purification-induced
population and coherence errors vanish in this limit within the
present effective noise model. This cancellation does not occur for
the non-default pure branch, for which $|z_{\bar d}|=1$. Thus, the
high-temperature suppression of the purification error is a
consequence of the fixed point of the depolarizing channel rather
than of the preparation scheme itself.

\subsubsection{Amplitude-damping (AD) channel}
\label{app:amplitude_damping}

As a representative nonunital noise mechanism, we next consider
amplitude damping toward the ground state,
\begin{equation}
    A_{p_G}(\rho)
    =
    K_0\rho K_0^{\dagger}
    +
    K_1\rho K_1^{\dagger},
\end{equation}
with the Kraus operators
\begin{equation}
K_0:=
\begin{pmatrix}
1 & 0\\
0 & \sqrt{1-p_G}
\end{pmatrix},
\qquad
K_1:=
\begin{pmatrix}
0 & \sqrt{p_G}\\
0 & 0
\end{pmatrix}.
\end{equation}
Taking the ground state to correspond to $z=+1$, an
amplitude-damping channel with effective damping probability $p_G$
acts on the longitudinal Bloch component as
\begin{equation}
    A_{p_G}:
    z
    \rightarrow
    (1-p_G)z+p_G.
    \label{eq:AD_bloch}
\end{equation}

For the purification circuit, successive applications of
Eq.~\eqref{eq:AD_bloch} for the $R$ and CNOT gates give
\begin{equation}
    \widetilde z_A^{\rm AD}
    =
    (1-p_{\rm CX})(1-p_{\rm R})z_A
    +(1-p_{\rm CX})p_{\rm R}
    +p_{\rm CX}.
\end{equation}
Thus, to first order in the effective error probabilities,
\begin{equation}
    \delta z_A^{\rm AD}
    :=\widetilde z_A^{\rm AD}-z_A
    \simeq
    (p_{\rm R}+p_{\rm CX})(1-z_A).
    \label{eq:delta_z_pur_AD}
\end{equation}

For the purification-based implementation, substituting
Eq.~\eqref{eq:delta_z_pur_AD} into the general population-error
formula, Eq.~\eqref{eq:pop_error_pur_general}, gives
\begin{equation}
    \Delta z_{n,\rm pur}^{\rm AD}
    \simeq
    (1-\widetilde{\gamma}^{\,n})
    (p_{\rm R}+p_{\rm CX})(1-z_A),
    \label{eq:pop_error_pur_AD}
\end{equation}
while substitution into the general coherence-error formula,
Eq.~\eqref{eq:coh_error_pur_general}, yields
\begin{equation}
    E_{n,\rm pur}^{\rm AD}
    \simeq
    \frac{n}{2}
    (p_{\rm R}+p_{\rm CX})(1-z_A)
    \left|
    \frac{
    \widetilde{\lambda}_\uparrow
    -
    \widetilde{\lambda}_\downarrow
    }{
    \widetilde{\lambda}_{\rm th}
    }
    \right|.
    \label{eq:coh_error_pur_AD}
\end{equation}

For the non-default branch of the post-processing reconstruction,
the $X$-gate error gives
\begin{equation}
    \delta z_{\bar d}^{\rm AD}
    \simeq
    p_{\rm X}(1-z_{\bar d}).
    \label{eq:delta_z_post_AD}
\end{equation}
For the hardware convention used in Fig.~\ref{fig:prep_error},
the ground state is the default state,
\begin{equation}
    z_d=+1,
    \qquad
    z_{\bar d}=-1,
    \label{eq:AD_default_convention}
\end{equation}
so that Eq.~\eqref{eq:delta_z_post_AD} reduces to
\begin{equation}
    \delta z_{\bar d}^{\rm AD}
    \simeq
    2p_{\rm X}.
    \label{eq:delta_z_post_AD_fig}
\end{equation}

Substituting Eq.~\eqref{eq:delta_z_post_AD} into the general
post-processing population-error formula,
Eq.~\eqref{eq:pop_error_post_general}, gives
\begin{equation}
    \Delta z_{n,\rm post}^{\rm AD}
    \simeq
    (1-\widetilde{\gamma}^{\,n})
    p_{\bar d}p_{\rm X}(1-z_{\bar d}),
    \label{eq:pop_error_post_AD}
\end{equation}
while substitution into the corresponding general coherence-error
formula, Eq.~\eqref{eq:coh_error_post_general}, gives
\begin{equation}
    E_{n,\rm post,\bar d}^{\rm AD}
    \simeq
    \frac{n}{2}
    p_{\rm X}(1-z_{\bar d})
    \left|
    \frac{
    \widetilde{\lambda}_\uparrow
    -
    \widetilde{\lambda}_\downarrow
    }{
    \widetilde{\lambda}_{\bar d}
    }
    \right|.
    \label{eq:coh_error_post_AD}
\end{equation}

For the default branch, no ancilla-preparation gate is required,
so that $\delta z_d=0$. The general result
Eq.~\eqref{eq:coh_error_default} therefore directly gives
\begin{equation}
    E_{n,\mathrm{post},d}^{\rm AD}=0.
    \label{eq:coh_error_default_AD}
\end{equation}

The temperature dependence of these expressions again reflects the
fixed point of the noise channel, but now in the opposite temperature
limit from the depolarizing case. Zero-temperature amplitude damping
has the ground state $z=+1$ as its fixed point. As $T\to0$, the thermal
ancilla approaches this state, $z_A\to1$, and hence
$\delta z_A^{\rm AD}\to0$. Accordingly,
Eqs.~\eqref{eq:pop_error_pur_AD} and
\eqref{eq:coh_error_pur_AD} show that both the purification-induced
population and coherence errors vanish in this limit within the
present effective amplitude-damping model.

For post-processing reconstruction, the non-default branch is not a
fixed point of amplitude damping under the convention
Eq.~\eqref{eq:AD_default_convention}. Its preparation-induced
population error nevertheless becomes irrelevant to the reconstructed
thermal population in the low-temperature limit because
$p_{\bar d}\to0$ in Eq.~\eqref{eq:pop_error_post_AD}. In contrast,
the coherence reconstruction may use the default branch alone, for
which Eq.~\eqref{eq:coh_error_default_AD} shows that the additional
preparation-gate contribution vanishes identically.

\subsubsection{Leading-order scaling}
\label{app:scaling}

The leading-order scalings can be read directly from the general
error expressions derived in Secs.~C.1 and C.2 together with the
polarization deviations evaluated for the specific noise models above.
For the population sector, Eqs.~\eqref{eq:pop_error_pur_general} and
\eqref{eq:pop_error_post_general} show that the preparation-induced
errors are linear in $\delta z_A$ and $\delta z_{\bar d}$, respectively.
For the coherence sector,
Eqs.~\eqref{eq:coh_error_pur_general} and
\eqref{eq:coh_error_post_general} show that the corresponding relative
errors are proportional to $n|\delta z_A|$ and
$n|\delta z_{\bar d}|$, while
Eq.~\eqref{eq:coh_error_default} gives zero for the default branch.

For both representative noise models considered above, the polarization deviations have the same leading orders in the preparation-gate error probabilities,
\[
    \delta z_A = O(p_R+p_{CX}),
    \qquad
    \delta z_{\bar d}=O(p_X),
\]
apart from noise-model- and temperature-dependent prefactors.
Specifically, these relations follow from
Eqs.~\eqref{eq:delta_z_pur_dep} and
\eqref{eq:delta_z_post_dep} for depolarizing noise, and from
Eqs.~\eqref{eq:delta_z_pur_AD} and
\eqref{eq:delta_z_post_AD} for amplitude damping.
Substitution into the general expressions therefore gives
\begin{equation}
\begin{aligned}
    \Delta z_{n,\rm pur}^{\rm prep}
    &= O(p_R+p_{CX}),\\
    \Delta z_{n,\rm post}^{\rm prep}
    &= O(p_X),\\
    E_{n,\rm pur}^{\rm prep}
    &= O\!\left[n(p_R+p_{CX})\right],\\
    E_{n,\rm post,\bar d}^{\rm prep}
    &= O(np_X),\\
    E_{n,\rm post,d}^{\rm prep}
    &=0.
\end{aligned}
\label{eq:summary_scaling_app}
\end{equation}
These are the scalings summarized in
Table~\ref{tab:error_summary}.

\end{document}